\RequirePackage{etoolbox}\csdef{input@path}{{style/}{graphics/}}
\documentclass[aoas,MSNbibl,nameyear,seceqn,dvips]{arximspdf}
\usepackage{multirow}
\usepackage{dcolumn}
\usepackage{graphicx}

\doi{10.1214/14-AOAS785} 
\volume{8}
\issue{4}
\pubyear{2014}
\firstpage{1966}
\lastpage{2001}
\docsubty{FLA}

\makeatletter
\newcolumntype{d}[1]{D{.}{.}{#1}}
\def\cov{\operatorname{cov}}
\makeatother

\begin{document}
\begin{frontmatter}

\title{Reconstructing past temperatures from natural proxies and
estimated climate forcings using short- and long-memory models}
\runtitle{Paleoclimate reconstruction}

\begin{aug}
\author[A]{\fnms{Luis}~\snm{Barboza}\corref{}\thanksref{M1,T1}\ead[label=e1]{luisalberto.barboza@ucr.ac.cr}},
\author[B]{\fnms{Bo}~\snm{Li}\thanksref{M2,T2}\ead[label=e2]{libo@illinois.edu}},
\author[C]{\fnms{Martin P.}~\snm{Tingley}\thanksref{M3,M31,T3}\ead[label=e3]{mpt14@psu.edu}}\\
\and
\author[D]{\fnms{Frederi G.}~\snm{Viens}\thanksref{M4,T4}\ead[label=e4]{viens@purdue.edu}}
\runauthor{Barboza, Li, Tingley and Viens}
\affiliation{Universidad de Costa Rica\thanksmark{M1},
University of Illinois at Urbana--Champaign\thanksmark{M2},
Pennsylvania State University\thanksmark{M3}, Harvard
University\thanksmark{M31} and
Purdue University\thanksmark{M4}}
\address[A]{L. Barboza\\
Centro de Investigaci\'on en Matem\'atica\\
\quad Pura y Aplicada (CIMPA)\\
Universidad de Costa Rica\\
2060 San Jos\'e\\
Costa Rica\\
\printead{e1}}
\address[B]{B. Li\\
Department of Statistics\\
University of Illinois at Urbana--Champaign\\
Champaign, Illinois 61820\\
USA\\
\printead{e2}}
\address[C]{M. P. Tingley\\
Department of Meteorology\\
\quad and Department of Statistics\\
Pennsylvania State University\\
State College, Pennsylvania 16801\\
USA\\
and\\
Department of Earth\\
\quad and Planetary Sciences\\
Harvard University\\
Cambridge, Massachusetts 02138\\
USA\\
\printead{e3}}
\address[D]{F. G. Viens\\
Department of Statistics\hspace*{78pt}\\
Purdue University\\
West Lafayette, Indiana 47907\\
USA\\
and\\
Department of Mathematics\\
Purdue University\\
West Lafayette, Indiana 47907\\
USA\\
\printead{e4}}
\end{aug}
\thankstext{T1}{Supported in part by NSF Grant DMS-09-07321.}
\thankstext{T2}{Supported in part by NSF Grants DMS-10-07686 and DPP-1418339.}
\thankstext{T3}{Supported in part by NSF Grant 1304309.}
\thankstext{T4}{Supported in part by NSF Grant DMS-14-07762 and by Government of Chile Conicyt MEC competition Grant 0011232.}

\received{\smonth{12} \syear{2012}}
\revised{\smonth{8} \syear{2014}}

%
\begin{abstract}
We produce new reconstructions of Northern Hemisphere annually averaged
temperature anomalies back to
1000~AD, and explore the effects of including external climate forcings
within the reconstruction and of
accounting for short-memory and long-memory features. Our
reconstructions are based on two linear models,
with the first linking the latent temperature series to three main
external forcings (solar irradiance,
greenhouse gas concentration and volcanism), and the second linking the
observed temperature proxy data (tree rings, sediment record, ice
cores, etc.) to the unobserved temperature series. Uncertainty is
captured with additive noise, and a rigorous statistical investigation
of the
correlation structure in the regression errors is conducted through
systematic comparisons between
reconstructions that assume no memory, short-memory autoregressive
models, and long-memory fractional Gaussian noise models.

We use Bayesian estimation to fit the model parameters and to perform
separate reconstructions of
land-only and combined land-and-marine temperature anomalies. For model
formulations that include forcings, both exploratory and Bayesian data
analysis provide evidence against models with no memory. Model
assessments indicate that models with no memory underestimate
uncertainty. However, no single line of evidence is sufficient to favor
short-memory models over long-memory ones, or to favor the opposite
choice. When forcings are not included, the long-memory models appear
to be necessary. While including external climate forcings
substantially improves the reconstruction, accurate reconstructions
that exclude these forcings are vital for testing the fidelity of
climate models used for future projections.

Finally, we use posterior samples of model parameters to arrive at an
\mbox{estimate} of the transient
climate response to greenhouse gas forcings of 2.5$^{\circ}$C (95\%
credible interval of
[$2.16$, $2.92$]$^{\circ}$C), which is on the high end of, but
consistent with, the
expert-assessment-based uncertainties given in the recent Fifth Assessment Report of the IPCC.
\end{abstract}

%
\end{frontmatter}


\setcounter{footnote}{3}

\section{Introduction}

An understanding of recently observed and projected future climate
changes [\citet{Alexander2013}] within the context of the natural
variability and dynamics of the climate system requires accurate and
precise reconstructions of past climate. As spatially wide-spread
instrumental temperature observations extend back to only about 1850,
it is necessary to turn to the noisy and sparsely distributed
paleoclimate record to characterize natural climate variability on
longer time scales. In addition, reconstructions of past climate allow
for important out-of-sample assessments of the Atmosphere--Ocean
General Circulation Models (GCM) that are used to project future
climate under various emissions scenarios [\citet
{ipcc5th05,ipcc5th09}]. While there is now a rich tradition of
inferring past climate from natural proxies, such as tree rings,
corals, ice cores, lake floor sediment cores and measurement on
speleothems [for recent reviews, see \citet{NRC06}; \citet
{jonesetal09,tingley2012}], many scientific and statistical challenges remain.

\subsection{Paleoclimatology context}

Reconstructions of past surface temperatures from networks of multiple
proxy types are prevalent in the climate science literature of the last
15 years---notable examples include \citet{Overpeck1997p3595},
Mann, Bradley and Hughes (\citeyear{mannbradley,mannbradley99}), \citet{luterbacher2004european}, \citet{moberg2005}, \citet{juckes},
Mann et~al. (\citeyear{mannzhang,mann2009global}), 
\citet{kaufman2009recent}, \citet{tingley2013Ext} and \citet{consortiumcontinental2013}.
While these studies have substantially increased our understanding of
past climate, limitations remain in terms of the statistical treatment
and uncertainty quantification. As described in \citet{tingley2012},
the most commonly used approaches to paleoclimate reconstruction are
all variants of multiple linear regression [see, e.g., Table~1
of \citet{christiansen2009}], regularized in some fashion to account
for the ``$p>n$'' problem in the estimation procedure. Examples of
particularly popular estimation approaches include regularized variants
of the Expectation--Maximization algorithm [\citet{dempster1977},
Mann et~al. (\citeyear{mann2007robust,mann2005testing}),
Rutherford et~al. (\citeyear{rutherford2003climate,bradley2005proxy}),
\citet{schneider2001analysis,zhang2004,steig2009}]
and principal component regression [\citet
{Cook1994,mannbradley,luterbacher2004european,Wahl2012}], which is
sometimes combined with canonical correlation analysis [\citet
{smerdon2010pseudoproxy}]. A common shortcoming of these studies lies
in the limited propagation of parameter uncertainty through the model,
including uncertainty in the estimation of regularization parameters;
for further discussion see \citet{schneider2001analysis},
\citet{smerdon2010pseudoproxy} and the supplement to \citet{Wahl2012}.

Recently, hierarchical modeling and Bayesian inference techniques have
been proposed and employed to reconstruct past climate from proxies\break 
[\citet{haslett}, \citet{boli1},
Tingley and Huybers (\citeyear{tingley1,tingley2,tingley2013Ext}),
\citet{werner2012pseudoproxy}].
Hierarchical modeling is a natural framework for including the
available scientific understanding of both the target climate process
(e.g., annual surface temperature anomalies) and how the various
natural proxies are causally affected by variations in the climate
system. Bayesian inference, in turn, provides a cohesive framework for
propagating uncertainty, while the posterior draws of the target
climate quantity are a more statistically precise and scientifically
useful result than a point estimate and associated uncertainty interval
[\citet{tingley2012}].

In this paper, we reconstruct Northern Hemisphere (NH) temperature
anomalies over the past millennium using a hierarchical Bayesian model
that describes temperature as linearly dependent on three important
climate forcings: green house gas concentrations, volcanic aerosol
concentrations and variations in solar irradiance. The proxies, in
turn, are modeled as linear in the latent temperature process.
Motivated by existing evidence of long-range correlation in temperature
series [e.g., \citet
{brody,benth,huybers2006links,Imbers}], we explore
the effects of specifying white noise (no memory), autoregressive
(short memory) and long-memory correlation structures for the two error
processes. To our knowledge, this is the first ensemble-based
paleoclimate reconstruction that includes the effects of climate
forcings, and the first systematic investigation of error structure in
the temperature reconstruction. As our method involves first reducing
the proxy data set to a single time series, and then inferring
hemispheric average temperature anomalies, rather than the spatial
pattern, our analysis is a form of composite-plus-scaling [\citet
{tingley2012}].

The external forcings used in the analysis are closely related to
global temperature evolution. The Intergovernmental Panel on Climate
Change (IPCC) has steadily increased its certainty level on stating the
causal relationship between increasing atmospheric concentrations of
anthropogenic greenhouse gases and increasing average global
temperatures, reaching the ``extremely likely'' level of 95\%
confidence in 2013
[\citet{ipcc5th10}]. The relationship between solar irradiance and
surface temperatures is studied in \citet{crowley96}, \citet{beer},
while \citet{briffajones}, \citet{crowley93vulc}, \citet
{crowley93} and
\citet{Landrum2013} analyzed the effect of volcanic activity on global
temperatures.

The conceptual study of \citet{boli1} demonstrated that temperature
reconstructions are improved when information about the climate forcing
is included in the reconstruction. We therefore explore the effects of
including these three major external forcings in our reconstructions,
reporting results for both cases. While the forcings are expected to
improve the reconstructions, we note that reconstructions that exclude
the forcings are necessary for the evaluations of GCMs [\citet
{ipcc5th05,ipcc5th09}] to avoid the circularity of using the same
forcings in the simulation of past climate and the reconstruction used
to assess the simulation.

\subsection{Long-memory modeling and estimation challenges}

To our knowledge, the error terms in all previous models for
multi-proxy climate reconstructions are assumed to be white or
autoregressive [AR; see, e.g., \citet{tingley2012}]. For instance,
\citet{boli1}, \citet{tingley1} and \citet{mcshane} use
AR(1) or AR(2)
errors, while reconstructions based on the Expectation--Maximization
algorithm or principal component \mbox{regression} have generally not
explicitly modeled temporal autocorrelation [see Section~8.7.4 of
\citet{tingley2012}].

The assessment of long-memory behavior in hierarchical models is
complicated by the fact that graphs of the autocorrelation and partial
autocorrelation functions (a.c.f. and p.a.c.f.) are generally not adequate
diagnostic tools. In addition, the short data streams we are faced with
disallow reliance on known asymptotic properties, while lack of
self-similarity means that inference on one range of frequencies cannot
apply to another. These issues are well known for widely used
long-memory time series models, such as fractional autoregressive integrated moving average (FARIMA) models [\citet{Beranbook}].
Misspecification of a long-memory process with a short-memory model can
lead to erroneously attributing long-memory effects to deterministic
trends or external forcings, and thus will affect uncertainty
quantification. Specifically, since long-memory models can exhibit
larger asymptotic variances than their relatively short-memory model
analogues [see \citet{chronopoulou2009variations} and references
therein], reported uncertainty levels under memory misspecification can
be lower than the nominal values.

Motivated by the limitations of the data, and our goal of using a
robust model, we focus on a simple long-memory model: linear regression
with fractional Gaussian noise (fGn) errors.
The theoretical question of estimating memory length for
nonself-similar models, such as
our hierarchical linear model, is notoriously difficult. Asymptotic
theory is still under development, and current work on high-frequency
or increasing-horizon versions of our model cannot yet be considered
definitive.
Online Supplement~\textup{A.1} in \citet{Barboza2014} provides
brief background information on long-memory estimation, while further
details can be found in references therein; see, in particular,
\citet{Gneiting2004}.

In the context of annual paleoclimate observations, time intervals
cannot be assumed small, and the calibration period is short. On
account of the long time intervals, we cannot use the local path
behavior of the data (e.g., H\"older continuity) as a proxy for long
memory---an approach that is possible for fGn-driven models where high
frequency data exists. Such models are asymptotically H\"
older-continuous in the limit of ultra-high frequency, with a single
parameter that also governs long memory.
On account of the short calibration period, methodologically sound
results from low frequency increasing-horizon asymptotics
[see \citet{tudorviensfbm}] cannot be used to measure long-range
dependence in our case, as there is simply not enough data. Instead we
resort to a fully Bayesian framework to estimate all parameters,
including those responsible for memory length, with the added benefit
of a complete evaluation and propagation of uncertainty.

This article is structured as follows. Section~\ref{Data} describes
the data sets used in the reconstruction, and Section~\ref{Model}
gives the details of the hierarchical Bayesian models. Section~\ref
{Results} presents the results of our Bayesian reconstructions,
including parameter posterior distributions and model validation
metrics; it compares models with different error structures and which
include or exclude the climate forcings. We also compare our results
with previous reconstructions and discuss the estimation of transient
climate response in Section~\ref{Comparisons} before summarizing our
quantitative conclusions and discussing remaining challenges in
Section~\ref{Conclusions}. Two online supplements provide further
details on the modeling framework and additional quantitative results
[see \citet{Barboza2014}].

\section{Data sets}\label{Data}

The analysis makes use of three distinct data sources: instrumentally
observed temperature anomalies (in $^{\circ}$Celsius) over the period
1900--1998; a~suite of temperature-sensitive proxies over the period
1000--1998 taken from the database originally described in \citet{mannzhang} and used additionally in \citet{mann2009global}; and
estimates of external climate forcings from 1000--1998~AD.

We make use of two different instrumental estimates of NH temperature
anomalies, both developed by the Climate Research Unit of the
University of East Anglia [\citet{brohan}]. The CRUTEM3v data set
(abbreviated hereafter as CRU) is an estimate of air surface
temperature anomalies over land, while HadCRUT3v (hereafter abbreviated
as HAD) is an estimate of combined land air- and marine sea-surface
temperatures. These data sets are widely used for the calibration of
proxy-based climate reconstructions [e.g., \citet
{mannzhang,luterbacher2004european,bradley2005proxy,kaufman2009recent,mcshane,tingley2013Ext}].
We make use of the variance-adjusted version of each data set to
facilitate comparisons with results from \citet{mannzhang}. While both
instrumental data sets extend back to 1850, we choose \mbox{1900--1998} as our
calibration period, as the sparsity of instrumental observations
results in less trustworthy hemispheric estimates prior to about 1900
[\citet{Smith10understandingsensitivities}].

The proxies used in our analysis are selected from the 1209
climate-sensitive proxies originally compiled in \citet{mannzhang}.\hskip.2pt\footnote{For more details on the data set, see the
NOAA-Paleoclimatology/World Data Center at \surl{http://www.ncdc.noaa.gov/paleo/pubs/pcn/pcn-proxy.html}.} This
compilation brings together a wide array of proxy types, including tree
ring widths and densities, marine sediment cores, speleothems (cave
deposits), lacustrine sediment cores, ice cores, coral records and
historical documentary information [see \citet{NRC06} and \citet
{jonesetal09} for further descriptions of each of these data types].
The proxy data are not raw observations, but are rather processed to
remove nonclimatic variability, such as age effects associated with
tree ring data. This type of processing results in a data product which
may be more directly interpreted as ``climate sensitive,'' according to
the paleoclimatology community. While it is common to base climate
reconstructions on the post-processed data, as is done here, we
acknowledge that doing so does neglect the uncertainty inherent in the
processing steps. We set aside for future research the challenge of
including the processing of raw climate proxy observations into
climate-sensitive series within the hierarchical framework developed
here. For further details concerning the processing of raw proxy
observations, see, for example, \citet{NRC06,jonesetal09}.

Estimates of the external climate forcings---atmospheric greenhouse
gas concentrations (C), solar irradiance (S) and volcanism (V)---are
described and plotted in \citet{boli1} and described more fully in
\citet{ammann2007solar}. The original greenhouse gas concentration time
series is in units of CO$_2$ equivalent in parts per million; the solar
irradiance series is in Watt${}/{}$m$^2$ and the
volcanic series is an estimate of the climate forcing, in $W/m^2$, derived sulphate measurements
on ice cores
[see \citet{ammann2007solar}
for further details].

\section{Model specification}\label{Model}

Hierarchical Bayesian models typically consist of three levels. The
data level describes the likelihood of the observations conditional on
a latent stochastic process. In our context, the latent process is the
time series of NH mean temperature anomalies, and the observations are
the proxies. The process level describes the parametric structure of
the latent process---often with recourse to prior scientific
information, such as knowledge of the underlying physical dynamics
[e.g., \citet{berliner}]. Finally, the prior level provides
closure and
allows for Bayesian inference by providing prior distributions for all
unknown parameters in the data- and process-levels. For a general
description of hierarchical modeling and Bayesian inference in the
paleoclimate context, see \citet{tingley2012}. Following
\citet{boli1},
the data-level models the proxies as a normal distribution with mean
equal to a linear function of the latent, unobserved true temperatures,
while the process-level models the latent temperature process as normal
with mean given by a linear function of the external forcings [\citet{boli1}]. We add to previous work by applying the model to actual
proxy data, as opposed to using pseudo proxy experiments derived from
climate model output [\citet{boli1}], as well as identifying
appropriate memory lengths in the error structures of the residuals at
both levels.

The Bayesian modeling framework is closely related to stochastic
filtering methods. An interesting application of classical Kalman
filtering [see \citet{kalman1961new}] to climatic reconstruction
is in
\citet{lee2008evaluation}, where the authors use forcings and a smaller
proxy data set to reconstruct temperatures on a decadal basis. However,
there are, to our knowledge, no practical tools for filtering with fGn
errors and, in addition, stochastic filters, which are adapted to
tracking moving signals dynamically in time, are notoriously poor at
estimating fixed parameters; see \citet{yang2008} and \citet{viens2}.
Thus, they are not an optimal choice for our exploration of short-
versus long-memory models in paleoclimate reconstructions. In contrast,
the Bayesian approach adopted here allows for all parameters to be
estimated simultaneously while avoiding the known estimation
difficulties inherent to filtering. Moreover, since the proxy
observations are not being updated over time, the sequential updating
property of filtering is not advantageous.

\subsection{Proxy data reduction}\label{rpconst}

It is desirable for several reasons to reduce the dimensionality of the
proxy data set, which consists of 1209 time series.
First, as there are only a limited number of years in the calibration
interval, dimension reduction can lead to a more parsimonious model,
avoid overfitting, and lead to more robust temperature reconstructions.
Second, our interest in inferring global mean temperatures rather than
spatial fields motivates a reduction, prior to fitting a hierarchical
model, to a single time series that reflects the shared variability
between the proxies that is likely attributable to a common, climatic origin.
Third, the proxy reduction is important in limiting the computational
burden of estimating parameters describing long memory; for a
comparison between computational and asymptotic efficiency for various
long-memory parameter estimators, see \citet{chronopoulou2009hurst}.
We therefore apply a sequence of steps to reduce the number of proxies
while attempting to retain as much climatically useful information as possible.

Following \citet{mannzhang}, we first select only those proxies that
are recorded at least as far back as 1000 AD and, in addition, have a
significant correlation with their closest instrumental time series
(marine or land) over their period of mutual overlap. We use local
temperature information in the screening procedure, as any proxy that
might correlate to hemispheric temperature with some degree of accuracy
should relate to its local temperature with higher precision
[\citet{mannzhang}]. Such a criterion does not take into account the
possibility of exploiting physical teleconnections that exist in the
actual climate system [\citet
{mannbradley,tingley2012,werner2012pseudoproxy}]. This screening
procedure yields 38 proxies whose distribution by type and location is
given in Table~\ref{table38}.
Tree rings represent the majority of proxies that pass the screening
criteria, consistent with the ubiquitous use of tree ring information
in annual resolution temperature reconstructions [\citet
{NRC06,jonesetal09,consortiumcontinental2013} and references therein].

%
\begin{table}
\tabcolsep=0pt
\caption{Geographical distribution of the 38 proxies by type}\label{table38}
\begin{tabular*}{\tablewidth}{@{\extracolsep{\fill}}ld{2.0}c@{}}
\hline
\textbf{Type} & \multicolumn{1}{c}{\textbf{\#}} & \textbf{Locations}\\
\hline
Tree ring & 16 & USA, Argentina, Norway, New Zealand, Poland, Sweden\\
Lacustrine & 7 & Mexico, Ecuador, Finland\\
Speleothem & 6 & China, Scotland, Yemen, Costa Rica, South Africa\\
Ice cores & 4 & Peru, Greenland, Canada\\
Other\tabnoteref{tt1} & 5 & China, Mongolia, Tasmania, New Zealand\\\hline
\end{tabular*}
\tabnotetext{tt1}{The category named ``Other'' contains data from composite temperature
reconstructions and historical documentary series.}
\end{table}

\begin{figure}

\includegraphics{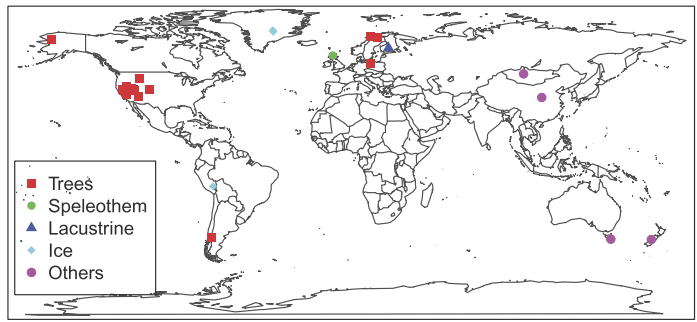}

\caption{Geographical distribution of the 25 proxy series.}\label{map25}
\end{figure}

A number of the 38 proxy series in Table~\ref{table38} show
undesirable properties given our assumption of a stationary
relationship between the proxies and temperatures. In particular,
several of the lacustrine and speleothem records feature much greater
variability in the early portion of the time interval than in the
calibration period. On such bases, we exclude 13 proxies, leaving a
total of 25; see Figure~B.1 and Table~B.1 in
the Online Supplement~\textup{B} for details [see \citet
{Barboza2014}]. The single lacustrine proxy included in the
reconstructions is the tiljander\_2003\_darksum series from Finland
[\citet{tiljander20033000}]. We apply a log-transformation on this
series in order to dampen the few years that feature very thick varves
[\citet{loso2009summer}], and to produce a series that is in-line with
the assumption of normal errors in our statistical models.
Figure~\ref{map25} shows the spatial locations of the 25 proxies.
%

To increase computational tractability, and to ensure that the
heterogenous spatial distribution of the proxies does not bias
estimates of the spatial average, we further reduce the 25 proxies into
a single series, termed the ``reduced proxy,'' via a weighted averaging
procedure. Intuitively, we seek a reduced proxy series that captures
the common signal of globally averaged climate reflected in the shared
variability between the proxies. We estimate the averaging weights used
to form the reduced proxy using least squares regression, first
centering and scaling each of the 25 proxy series over the period
1000--1998. Denoting these scaled proxies as $P_{i,t},i=1,\ldots,25$
and $t=1000,\ldots,1998$ and the HAD or CRU series as~$T_{t}$ (mean
temperature anomalies), we estimate the weights via an ordinary least
squares fit to $T_{t}=a_{0}+\sum_{i=1}^{25}a_{i}P_{i,t}+\varepsilon
_{t}$, where $\varepsilon_{t}$ is white noise.
Since most of the proxies end after 1982, here we fit the model using
only the data from 1900 to 1982.
The least squares parameter estimates $\widehat{a_{0}},\ldots,\widehat
{a_{25}}$ provide a weighted average of proxies that
maximizes the explained variance. Denote the reduced proxy as $\mathrm{RP}_{t}$, then
%
%
\begin{equation}
\mathrm{RP}_{t}=\widehat{a_{0}}+\sum_{i=1}^{25}
\widehat{a_{i}}P_{i,t}. \label{reducedproxdef}%
\end{equation}

The percentage of variation in temperatures that can be explained by
the reduced proxy is $R^{2}=77.48\%$ for the HAD data set and
$R^{2}=58.25\%$ for the CRU data set; note that the $R^2$ is higher for
the HAD data set despite all proxies being terrestrial. The proxies are
selected on the basis of local correlations, and the higher percentage
of explained variation with the HAD data set is indicative of the fact
that temperature observations at the locations of the proxies (many of
which are coastal) are better at predicting global land and sea
temperatures than global land-only temperatures.
Note that colinearity is not an issue, as the $P_{i,t}$ do not feature
strong correlations with one another, and, in addition, our interest
lies in the linear combination of $P_{i,t}$ rather than the
coefficients $\widehat{a_{i}}$.

The geophysical distribution of the weights (in percentage of absolute
value) is displayed in Tables~B.3 and B.4 in
Online Supplement~\textup{B} of \citet{Barboza2014}. For both
HAD and CRU data sets, proxies in the United States are most heavily
weighted, followed by the Mongolian composite. The remaining countries
have a fairly uniform distribution, with no single country exceeding
the 8\% level (HAD) or 7\% level (CRU). Our selected proxies therefore
have broad spatial coverage, inasmuch as possible with the available
data. The weights heavily concentrate on the ``Tree rings'' and
``Other'' categories, consistent once more with the prevalence of tree
ring series in climate reconstructions [e.g., \citet
{Overpeck1997p3595,mannbradley,luterbacher2004european,moberg2005,tingley2013Ext,consortiumcontinental2013}].
The weight for the single lacustine series, from \citet
{tiljander20033000}, is less than 8\% for both HAD and CRU data sets,
indicating that it exerts a limited control on the overall
reconstructions. The limited influence of this lacustrine series is of
particular importance given the known difficulties in calibrating it,
due to the potential of anthropogenic impact on the lake catchment
[\citet{tiljander20033000,Mannetal2008supp}]; we return to this point
in Section~\ref{TRresult}.

The modeling approach taken here, based on a weighted average of
proxies that pass a local screening condition, does not explicitly
consider long-range spatial dependencies, or teleconnections, within
the climate system. Another option would be to set the reduced proxy to
the leading principal component of the 25 proxies that pass the
screening test. Such an approach would extract the dominant common
signal shared by the proxies, whereas for the purposes of this analysis
we are more interested in retaining the common temperature signal they
share. While methods based on principal component or canonical
correlation analysis are prevalent in paleoclimatology, both for the
reconstruction of spatial patterns and (as here) spatial averages,
there is ongoing debate as to the merits of such methods; see
\citet
{Cook1994,NRC06,Wahl2012,tingley2012,werner2012pseudoproxy,consortiumcontinental2013}
for discussion.

\subsection{Examination of long-memory correlation in the proxy data}\label{Proxy}

While the temperature--proxy relationship is almost universally assumed
to be linear [e.g., \citet{luterbacher2004european}, \citet
{bradley2005proxy}, \citet{boli1}, \citet{tingley2},
\citet{kaufman2009recent}, \citet{mcshane}, \citet
{christiansen2011}, \citet{smerdon2010pseudoproxy}, and each of
the methods in Table~1 of \citet{christiansen2009} and discussed
in Section~8 of \citet{tingley2012}],
the correlation structure in the error term has not been thoroughly
studied. The choice of model for the correlation structure is of
particular importance, as its adequacy directly affects the accuracy
and precision of the uncertainty quantification associated with the
reconstruction.
Here we consider models of the form
%
%
\begin{equation}
\mathrm{RP}_{t}=\alpha_{0}+\alpha_{1}T_{t}+
\sigma_{p}\eta_{t}, \label{eq2}%
\end{equation}
where $\eta_{t}$ is a zero-mean, unit-variance stationary stochastic
process and $\sigma_{p}$ a constant variance parameter. We fit model
(\ref{eq2}) using least-squares over the 1900--1982 interval, using
either the HAD or CRU as $T_t$, and examine the correlation structure
of the resulting residuals.

We first explore the correlation structure of $\eta_{t}$ using
estimates of the spectral density, $f(\lambda)$, of the empirical
residuals. If the residuals have long-memory behavior, then the
logarithm of the spectrum will feature a negative slope with respect to
log-frequency. More specifically, a stationary stochastic process
$X_{t}$ is generally said to have long memory when its autocovariance
function $\gamma(n):=\cov( X_{t+n},X_{t} )$ decays at the rate
$n^{2H-2}$ for large time lag $n$, where $0.5 < H < 1$ is the
long-memory parameter.
This behavior is essentially equivalent to requiring that $f(\lambda)$
have a singular behavior $\lambda^{1-2H}$ for small frequencies
$\lambda$ [see \citet{Beranbook}]. Since $1-2H<0$ for long-memory
models, the plot of $\log f(\lambda)$ against $\log\lambda$ for a
long-memory model will be approximately a straight line with negative
slope $1-2H$. While spectral methods are not generally accepted as a
formal way to estimate $H$, save for very simple models, they do offer
a useful diagnostic tool to evaluate the long-memory structure in the
data [see \citet{Beranbook}].

\begin{figure}
\begin{tabular}{@{}cc@{}}

\includegraphics{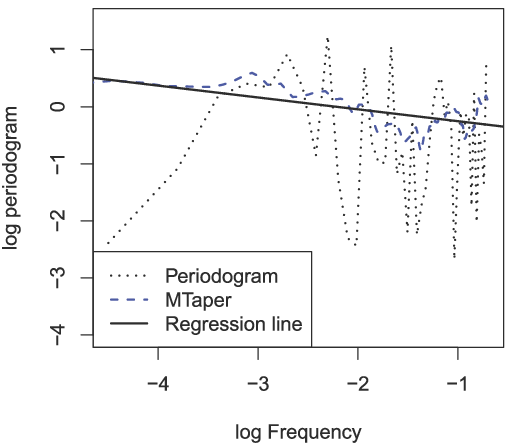}
& \includegraphics{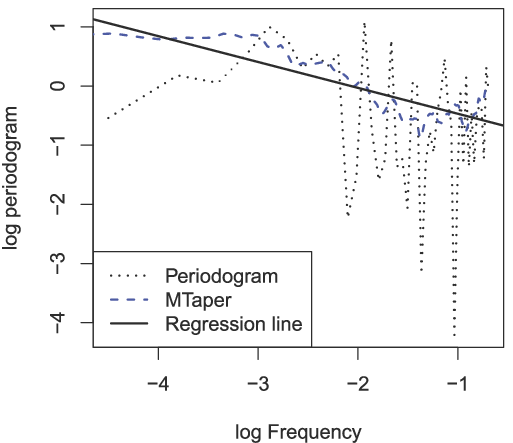}\\
\footnotesize{(a) HAD data set}&\footnotesize{(b) CRU data set}
\end{tabular}
\caption{Spectral estimates on a log--log scale, with frequency units
of cycles per year. The regression line is computed by regressing the
log of multitaper estimator onto the log-frequencies.}\label{fig3}
\end{figure}

Based on the regression residuals from equation (\ref{eq2}), we
compute two widely used estimators of the spectral density: the
periodogram and the adaptive multitaper estimator [see Online
Supplement~\textup{A} in \citet{Barboza2014} for a brief
description for each estimator]. Figure~\ref{fig3} shows both
estimators on a log--log scale for the HAD and CRU data sets, respectively.
In both cases, the multitaper spectral estimator features a clear
negative slope on the log--log scale, indicating possible long-memory
behaviors. Results for the periodogram are less striking than the
multitaper estimate, but still show a negative slope in log--log space.

To examine more formally the long memory behavior of the residuals, we
employ the test developed by \citet{robinson} (Section~\ref{othertest}
presents results of alternative tests). To introduce the idea of this
method briefly, consider a stationary process $X_{t}$ with spectral
density $f(\lambda)$. The $f(\lambda)$ may satisfy the power law
$f(\lambda)\sim G\lambda^{1-2H}$ as $\lambda\rightarrow0$ for a
positive value $G$ and some $H\in(0,1)$. The so-called Hurst parameter
$H$ measures the length of the correlation as illustrated by the
negative slope of the spectrum in Figure~\ref{fig3}. Typical examples
that follow this power law include FARIMA and fGn.
This fGn is the discrete-time stationary Gaussian process that is the
first-order difference process of the so-called fractional Brownian
motion (fBm) process evaluated at integer times. The spectrum of the
distributional derivative of the fBm process is proportional to
$\lambda^{1-2H}$. The spectrum of fGn has the same behavior
asymptotically for small $\lambda$.
Historically, the parameter $H$ first made its appearance when fBm was
introduced by \citet{kolmogorov1940wienersche}; the name \emph{Hurst}
arose after Mandelbrot proposed that fBm might be a good model to
explain the power behavior of a statistic introduced by the hydrologist
H.~E.~Hurst to study yearly levels of the Nile river; see \citet
{mandelbrot1965classe,mandelbrot} and the account in \citet
{taqqu2013benoit}.
More information on fGn can be found in Online Supplement~\textup{A} in \citet{Barboza2014}. The FARIMA model depends on a
parameter usually denoted by $d=H-1/2$ and features a spectral density
with the same low-frequency and long-memory asymptotics as fGn.

%

The null hypothesis for the \citet{robinson} test is $H=0.5$ (no
memory), while the alternative hypothesis is $H>0.5$ (long-memory). The
test is based on the asymptotic normality of the semiparametric
Gaussian estimate of $H$. Other tests for the memory length are
reviewed in \citet{murphy}, who recommend Robinson's test due to its
power properties and its good performance for relatively small samples
when combined with bootstrap resampling.

We perform Robinson's test on the regression residuals in (\ref{eq2}),
resulting in $p$-values of $0.0258$ for HAD and $0.0002$ for CRU. Both
data sets therefore show strong evidence, according to Robinson's test,
in favor of rejecting the null hypothesis of $H=0.5$. Note that, the
test, while consistent with long memory, does not provide evidence in
favor of long-memory correlations over shorter nonzero ones; in the
model-comparison exercises below (Section~\ref
{secvalidation-measures}), we also consider models which contain short
memory, AR(1) errors.

\subsection{Examination of long-memory behavior in the temperature
anomalies} \label{Temp}

In the specification of the process level of the hierarchical model, we
follow \citet{boli1} and model the latent temperature as linear in the
external forcings. We apply the following transformations to the
forcings, where $S$, $V$ and $C$ are, respectively, the time series of
solar irradiance, volcanism and greenhouse gases:
\begin{itemize}
\item$\tilde V_t=\log{(-V_t+1)}$. Exploratory data analysis indicated
that this transformation increases the explanatory power of volcanism.
From a physical standpoint, it dampens the effects of very large
events, and thus provides a form of regularization given the larger
uncertainties associated with the larger $V$ values [\citet{boli1}].
\item$\tilde C_t=\log(C_t)$. Following \citet{ipcc9}, we use a
log-transformation to approximate the radiative forcing due to changes
in the equivalent CO$_2$ concentration.
\end{itemize}
The resulting process-level model is
%
%
\begin{equation}
T_{t}=\beta_{0}+\beta_{1}S_{t}+
\beta_{2}\tilde V_{t}+\beta_{3}\tilde
C_{t}+\sigma_{T}\varepsilon_{t}, \label{eq1}
\end{equation}
where $\varepsilon_{t}$ denotes a stationary stochastic process with zero
mean and unit variance, and $\sigma_{T}$ is a constant variance
parameter. \citet{boli1} employ an AR(2) for the error term, based on
an examination of auto- and partial auto-correlation functions.
However, in a similar situation,
\citet{Beranbook} shows that the residuals are appropriately
modeled as
FARIMA (0, $d=0.4$, 0), with Hurst parameter $H=d+0.5=0.9$. \citet
{benth} and \citet{brody} also provide examples of estimation of
long-memory parameters over regression residuals on temperature series
for specific locations in Norway and England, respectively,
while \citet{huybers2006links} provides statistical evidence of a
power-law behavior in the spectrum of surface temperatures. Finally,
\citet{Imbers} use a long-memory fractional-differencing process that
is very similar to fGn in terms of its asymptotic long-memory behavior,
in order to test the presence of an anthropogenic impact on present-day
temperatures.

\begin{figure}
\begin{tabular}{@{}cc@{}}

\includegraphics{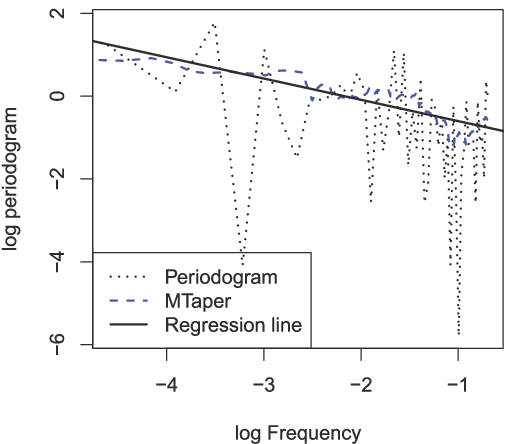}
& \includegraphics{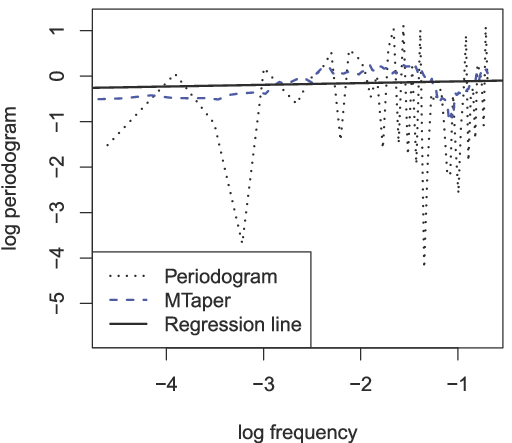}\\
\footnotesize{(a) HAD data set}&\footnotesize{(b) CRU data set}
\end{tabular}
\caption{Spectral estimates in log--log scale. The regression line is
computed by regressing the logarithm of multitaper estimator on
log-frequencies. The frequency units are cycle${}/{}$year.}\label{fig2}
\end{figure}

We repeat the same diagnostic procedure and hypothesis testing as in
Section~\ref{Proxy} to assess the long memory behavior of $\varepsilon
_{t}$. We first fit model (\ref{eq1}) using the ordinary least-squares
criterion, and find $R^{2}$ values of 73\% for HAD and 66\% for CRU,
indicating the strong explanatory power of the forcings.
Figure~\ref{fig2} plots spectral density estimates in log--log space,
for both HAD and CRU, and shows that HAD, but not CRU, exhibits a
negative slope.
%
The $p$-value associated with Robinson's test is $8.39\times10^{-7}$ for
HAD and $0.058$ for CRU, indicating strong evidence against no memory
for the HAD data set, but not for the CRU data set.

As there is value in reconstructions that exclude the forcings (e.g.,
for the purpose of General Circulation Model assessment), we also
consider a reduced form of the process-level model that exclude the
forcings, and models climate variability as a purely stochastic
process. Applying Robinson's test to the CRU and HAD data sets results
in $p$-values of $2.12\times10^{-10}$ for both cases, where we can note
that the amount of evidence against no memory increases when we exclude
the forcings.

\subsection{Other tests}\label{othertest}

We briefly discuss results for several alternatives to Robinson's test.
Beran's test [see \citet{Beran2}] evaluates the goodness of fit of a
particular stochastic process model (e.g., fGn) to a realization of a
time series. Let $X_{t}$ be a stationary Gaussian process with spectral
density $f(\lambda)$, whose realization one observes. When testing for
fGn, for instance, if $f(\lambda,H)$ is the spectral density of an fGn
process with Hurst parameter $H$, then the null hypothesis for Beran's
test is $H_{0}\dvtx f(\lambda)=f(\lambda,H)$ and the alternative is
$H_{a}\dvtx f(\lambda)\neq f(\lambda,H)$. Both the Robinson and Beran
tests base their test statistics on the Whittle estimator of $H$, which
enjoys the desirable property of insensitivity to certain changes of
scale [see Online Supplement~\textup{A} in \citet{Barboza2014}
for additional technical details].

We performed Beran's test on six data sets: the four residuals from the
HAD and CRU data sets, for both the proxy [equation~(\ref{eq2})] and
instrumental [equation~(\ref{eq1})] equations, and the two HAD and CRU
temperature data series themselves with no forcings. To test the
presence of memory, we use three distinct memory structures: fractional
Gaussian noise, AR(1) and AR(2). The null hypothesis in each test is
that the data comes from a spectral density equal to that of the given
memory structure. Thus, a nonrejection of the null is not inconsistent
with the tested memory structure. For our eighteen Beran's tests, the
corresponding $p$-values are shown in Table~\ref{tabberantests}.
The results indicate that Beran's test cannot reject the null in any of
the eighteen cases; this is consistent with the presence of memory, but
the tests do not point to a preferred memory structure.

%
\begin{table}
\tabcolsep=0pt
\caption{Results of Beran's test applied to the residuals from the HAD
and CRU data sets, for both the proxy [equation~(\protect\ref{eq2})]
and instrumental [equation~(\protect\ref{eq1})] equations, with or
without forcings, under three null hypotheses}\label{tabberantests}
\begin{tabular*}{\tablewidth}{@{\extracolsep{\fill}}lccc@{}}
\hline
\textbf{Model} & \textbf{fGn} & \textbf{AR(1)} & \textbf{AR(2)} \\
\hline
HAD-proxy & 0.77 & 0.40 & 0.76 \\
CRU-proxy & 0.91 & 0.72 & 0.92 \\
HAD-temp. & 0.56 & 0.58 & 0.59 \\
CRU-temp. & 0.73 & 0.33 & 0.40 \\
HAD-temp. (\textit{no forcings}) & 0.61 & 0.63 & 0.67 \\
CRU-temp. (\textit{no forcings}) & 0.46 & 0.19 & 0.47 \\
\hline
\end{tabular*}
\end{table}

Finally, we apply the test proposed by \citet{Davies1987}; see
Section~A.1.2  for technical details. The fGn is used as
the underlying parametric model for this test, and the null and
alternative hypotheses are identical to Robinson's test: $H=0.5$ (no
memory) versus $H>0.5$ (long-memory). Thus, in contrast with Beran's
test, rejection of the null is evidence against no memory. As in
Beran's test, we use the four residuals from the HAD and CRU data sets,
for both the proxy [equation~(\ref{eq2})] and instrumental
[equation~(\ref{eq1})] equations, and the two HAD and CRU series with no
forcings. $P$-values in Table~\ref{tabdaviesharte}
show that the null can be rejected in three out of four cases when we
include forcings within the models, and in the two cases without
forcings. In fact, the evidence against no memory increases when we
exclude forcings.

%
\begin{table}[b]
\tabcolsep=0pt
\tablewidth=250pt
\caption{Results of Davies and Harte's test applied to the residuals
from the HAD and CRU data sets, for both the proxy [equation~(\protect\ref{eq2})] and instrumental [equation~(\protect\ref{eq1})] equations,
with or without forcings, under the null hypothesis of no memory}\label{tabdaviesharte}
\begin{tabular*}{\tablewidth}{@{\extracolsep{\fill}}lc@{}}
\hline
\textbf{Model} & \textbf{Davies \& Harte} \\
\hline
HAD-proxy & 0.046 \\
CRU-proxy & 0.000 \\
HAD-temp. & 0.010\\
CRU-temp. & 0.436\\
HAD-temp. (\textit{no forcings}) & 0.000 \\
CRU-temp. (\textit{no forcings}) & 0.000 \\
\hline
\end{tabular*}
\end{table}

No single method employed here is a perfect indicator for the presence
or absence of memory in our error processes. Taken together, however,
the spectral density estimates and applications of the tests of
\citet{robinson}, \citet{Beran2} and \citet{Davies1987}
indicate to us that
the possibility of memory, long or short, cannot be ignored in
developing models for the residuals or for the HAD and CRU series
themselves. In Section~\ref{Results}, we further investigate the
memory properties of the residual processes, via Bayesian parameter
estimates and reconstruction validation measures.

\subsection{Hierarchical Bayesian model with long- or short-memory errors}\label{HBmLMe}
Gi\-ven the statistical evidence for long- or short-memory correlation in
the empirical residuals from equations (\ref{eq2}) and (\ref{eq1}),
and the implication for fGn or AR model by Beran's test, we explore the
results of modeling the errors using either fGn or AR processes. As the
strategy for fitting the hierarchical Bayesian reconstruction is
similar in each case, we present details for the more computationally
involved fGn error assumption. Comparisons between various modeling
choices (long memory vs. short memory vs. no memory; with or without
forcings) are given in Section~\ref{TRresult}. A summary of the data
and process levels of the hierarchical model is as follows:
%
%
\begin{eqnarray}
\label{eqprincipal} \mathrm{RP}_{t} & =&\alpha_{0}+\alpha_{1}T_{t}+
\sigma_{P}\eta_{t},
\nonumber
\\[-8pt]
\\[-8pt]
\nonumber
T_{t} & =&\beta_{0}+\beta_{1}
S_{t}+\beta_{2}\tilde V_{t}+\beta_{3}
\tilde C_{t}+\sigma_{T}\varepsilon_{t},
\end{eqnarray}
where $\eta_{t}$ and $\varepsilon_t$ are independent fGn processes with
respective parameters $H\in(0,1)$ and $K \in(0,1)$ which control the
long-memory behavior. We assume these models hold throughout the entire
prediction period (1000--1899) and calibration period (1900--1998).
Independence between $\varepsilon_{t}$ and $\eta_{t}$ is a reasonable
assumption, as $\eta_{t}$ represents the stochastic aspect of the
proxies that is not explained by the climate, while $\varepsilon_{t}$ is
the long-memory aspect of the climate not attributable to the forcings.

The modeling framework [equation~(\ref{eqprincipal})] is based on the
assumption that the relationship between the proxies and temperatures
is invariant through time. While stationarity may be an idealized
assumption, we note that our data selection procedure ensures that
stationarity is at the very least not an unreasonable assumption, while
the short calibration period precludes a more in-depth study of
possible nonstationarity in the temperature--proxy relationship.
Moreover, we note that the modeling framework could be made more
realistic by specifying a (possibly independent) error structure for
each individual proxy series. We do not pursue these specifics here,
but rather focus on exploring the effects of long memory and forcings
on the reconstruction.

Following \citet{boli1}, we define\vspace*{1pt} the following prior distributions
for the parameters $\bolds{\alpha}:=(\alpha_{0},\alpha
_{1})^{\mathrm{T}}$, $\bolds{\beta}:=(\beta_{0},\beta
_{1},\beta
_{2},\beta_{3})^{\mathrm{T}}$, $\sigma_{1}^{2}$, $\sigma_{2}^{2}$,
$H$ and~$K$:
\begin{itemize}
\item$\bolds{\alpha}\sim N((0,1)^{\mathrm{T}},\mathbf{I}_{2})$;
$\bolds{\beta}\sim N((0,1,1,1)^{\mathrm{T}},\mathbf{I}_{4})$;\vspace*{1pt}
\item$\sigma_{T}^{2} \sim \operatorname{IG}(2;0.1)$, $\sigma_{P}^{2} \sim \operatorname{IG}(2;0.1)$;\vspace*{1pt}
\item$H\sim\operatorname{Unif}(0,1)$; $K\sim\operatorname{Unif}(0,1)$;
\end{itemize}
where $\mathbf{I}_{n}$ is the identity matrix of dimension $n$.

Let $\mathbf{T}_{u}=(T_{1000},\ldots,T_{1899})$ denote the vector of
unknown temperatures and $\mathbf{T}_{0}=(T_{1900},\ldots,T_{1998})$
the vector of instrumental temperatures. Our goal is to infer $\mathbf
{T}_{u}$ based on $\mathbf{T}_{0}$, $\mathrm{RP}$, $S$, $\tilde V$ and $\tilde
C$. The full conditional posterior distributions of $\mathbf{T}_{u}$
and all unknown parameters save $H$ and $K$ can be derived explicitly,
thus allowing for standard Gibbs sampling in the Markov chain Monte
Carlo (MCMC) method. We resort to Metropolis--Hasting steps to sample
$H$ and $K$. The derivation of full conditional distributions can be
found in Online Supplement~\textup{A} in \citet{Barboza2014}.
We implement the MCMC using a number of \textsf{R} packages:
\mbox{\texttt{MCMCpack}} [\citet{MCMCpack}], \texttt{mvtnorm} [\citet
{mvtnorm}], \texttt{ltsa} [\citet{ltsa}] and \texttt{msm} [\citet{msm}].

\section{Numerical results}\label{Results}

The diagnostic tests in Section~\ref{Model}, while providing no
conclusive evidence for the presence of long or short memory, indicate
the possibility of certain correlations. In order to further
investigate appropriate models for error structures and to assess the
benefit of incorporating external forcings in the reconstruction, we
compare eight model variants on the basis of their parameter estimates
and reconstruction validation metrics:
\begin{longlist}[B]
\item[A:] Possible long memory ($H$ and $K$ not fixed), with external forcings.
\item[B:] Possible long-memory error in (\ref{eq2}) and AR(1) error
in (\ref{eq1}), with external forcings.
\item[C:] AR(1) error in (\ref{eq2}) and possible long-memory error
in (\ref{eq1}), with external forcings.
\item[D:] AR(1) errors in (\ref{eqprincipal}), with external forcings.
\item[E:] No memory ($H=K=\frac{1} 2$), with external forcings.
\item[F:] Possible long memory ($H$ and $K$ not fixed), no external
forcings ($\beta_i=0, i=1,2,3$).
\item[G:] AR(1) errors in (\ref{eqprincipal}), no external forcings
($\beta_i=0, i=1,2,3$).
\item[H:] No memory ($H=K=\frac{1} 2$), no external forcings ($\beta
_i=0, i=1,2,3$).
\end{longlist}
The AR(1) model is included, as it features short memory---an
intermediate model between assuming fGn and assuming uncorrelated white
noise. We refer to Scenarios~\textup{E} and~\textup{H} as having no memory, as they are
based on Gaussian white noise errors that are independent and thus have
no memory. Scenario \textup{B} allows for a long-memory model for the proxies
while assuming short memory in the temperature residuals, while
Scenario~\textup{C} reverses the assumptions of Scenario~\textup{B}.

For reconstructions using both the HAD and CRU instrumental records, we
sample 5000 times from the posterior distribution and discard the first
1000 replicates to account for the burn-in period. The details of
posterior samples are shown in Online Supplement~\textup{B}
[see \citet{Barboza2014}]. Here we summarize the results and show a
selection of representative plots and focus on reconstructions using
the HAD data set.

\subsection{Bayesian parameter estimates}\label{secBayesian}
\begin{figure}
\begin{tabular}{@{}cc@{}}

\includegraphics{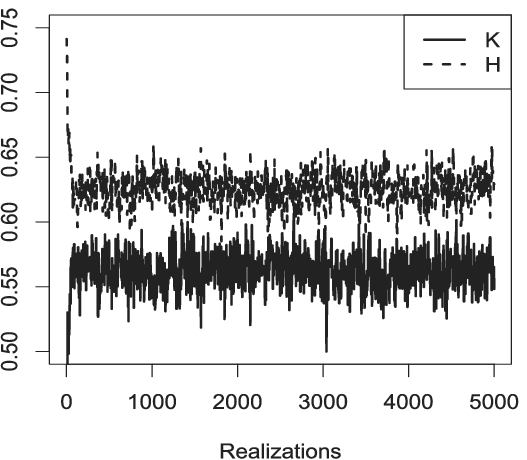}
&
\includegraphics{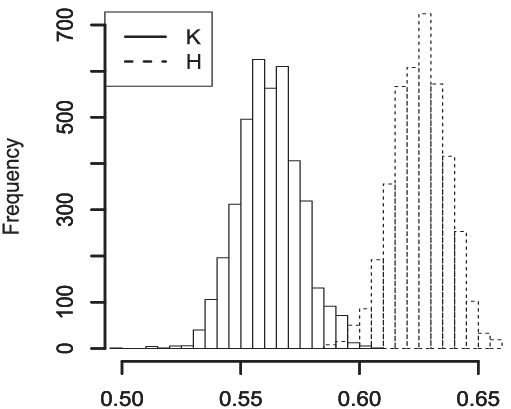}
\\
\footnotesize{(a) Traceplots}&\footnotesize{(b) Histograms}
\end{tabular}
\caption{Bayesian estimation of $H$ and $K$ based on HAD data set, Scenario~\textup{A}.}\label{HestHAD}
\end{figure}

\begin{figure}
\begin{tabular}{@{}cc@{}}

\includegraphics{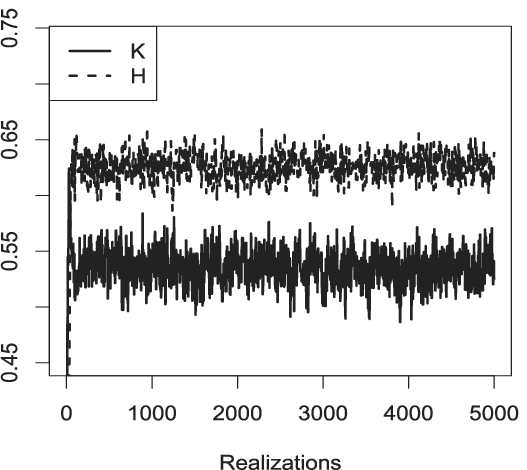}
&
\includegraphics{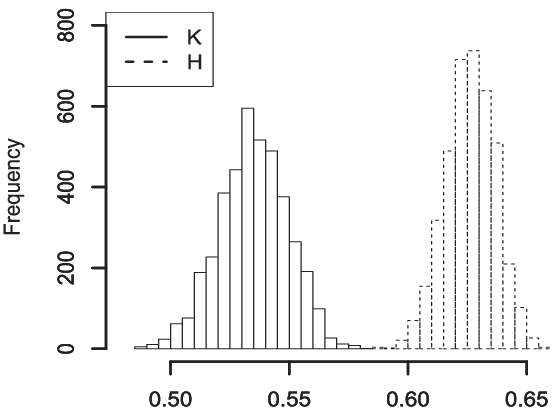}
\\
\footnotesize{(a) Traceplots}&\footnotesize{(b) Histograms}
\end{tabular}
\caption{Bayesian estimation of $H$ and $K$ based on CRU data set,
Scenario \textup{A}.}\label{HestCRU}
\end{figure}

We first examine parameter estimates using the HAD data set and
including the forcings.
Figure~\ref{HestHAD} shows trace plots and histograms of the $H$ and
$K$ parameters that are responsible for long memory in Scenario~\textup{A}.
Visually, the posterior draws quickly stabilize; see Section~4.3 for a
formal assessment of convergence for these and other parameters.
The histograms of $H$ and $K$ for the HAD reconstruction clearly
indicate that both parameters are significantly greater than 0.5,
suggesting that the data are consistent with a long-range correlation
model. Figure~\ref{HestCRU} shows the posterior distribution of $H$
and $K$ for the CRU reconstruction. The distribution of $H$ (memory
structure of the proxy residuals) is similar to that arising from the
HAD analysis, whereas the posterior distribution of $K$ for the CRU
analysis is centered on smaller values than for HAD, but still remains
significantly greater than 0.5. The larger value of $K$ for the HAD
data set, which includes the oceans, is in line with intuition, on
account of the larger heat capacity of the oceans resulting in a longer
timescale response to changes in the forcings.

%
%

%

For Scenarios \textup{B}, \textup{C} and \textup{D}, both HAD and CRU show that all AR(1)
parameter estimates are significantly greater than zero, and all
long-memory parameter estimates are significantly greater than 0.5 [see
Figures~B.2, B.3 and B.4  in the Online Supplement~\textup{B} of
\citet{Barboza2014}]. For Scenarios \textup{F} and \textup{G}, which exclude the
forcings, and
have, respectively, long and short memory, Bayesian posteriors for the
memory parameters provide evidence against models with no memory at
higher levels of certainty than for models that include forcings,
especially in the CRU case; see Figure~B.5 in Online
Supplement~\textup{B} [see \citet{Barboza2014}]. These results
indicate that while there is a certain amount of memory in the error
structures, there is insufficient evidence to select between short- or
long-memory assumptions. In the subsequent section, we resort to
reconstruction validation metrics to compare different models.

Posterior samples for the process-level regression coefficients (the
$\beta_i$) for Scenario \textup{A} show that the transformed volcanic and
greenhouse gas forcing series are meaningful predictors of the
temperature evolution for both HAD and CRU, while solar irradiance is
less influential (Figures~B.8 and B.12).
While the forcings are useful predictors of past temperatures, we
stress that the reconstructions that exclude the forcings are also of
scientific interest. Such reconstructions may not provide the most
accurate estimates of past climate fluctuations, but provide necessary
test beds for assessing the GCMs used to project future climate, since
comparisons between forcings-based reconstructions and GCM simulations,
which are based on the same forcings, would pose circularity issues.

\subsection{Validation measures}\label{secvalidation-measures}

We provide quantitative assessments of the eight reconstructions using
a number of statistical measures: squared bias (squared sample mean of
differences between the posterior mean and the observed anomalies);
variance (sample variance of the differences used in bias calculation);
root mean squared error (RMSE); empirical coverage probabilities (ECP)
of the credible intervals at the $95\%$ and $80\%$ levels; Interval
Scores (IS) at the $95\%$ and $80\%$ levels; and, since we obtain MCMC
samples from the predictive distribution, the Continuous Ranked
Probability Score (CRPS). The ECP measures the accuracy of the
uncertainty quantification and values closer to nominal level are more
desirable, while the IS and CRPS provide more nuanced assessments of
the posterior predictive distributions, rewarding both calibration and
sharpness simultaneously; details of these scoring rules are available
in \citet{Gneiting2007a,Gneiting2007,Gschlossl2007} and Online
Supplement~\textup{A.3} in \citet{Barboza2014}.
For convenience, we report the negative IS and CRPS so that smaller
values indicate higher quality predictions.

%
\begin{table}
\tabcolsep=0pt
\caption{Validation measures for the eight reconstruction scenarios,
using both HAD and CRU data sets. Scenarios \textup{F}, \textup{G} and \textup{H}, which include
no forcings, are italicized in this table}\label{validateHAD}
\begin{tabular*}{\tablewidth}{@{\extracolsep{\fill}}@{}lccccccccc@{}}
\hline
& \textbf{Scenarios} & \textbf{Sq. bias} & \textbf{Variance} & \textbf{RMSE} & \textbf{ECP}$\bolds{_{95}}$
& \textbf{ECP}$\bolds{_{80}}$ & \textbf{I}$\bolds{_{95}}$ & \textbf{IS}$\bolds{_{80}}$ &  \textbf{CRPS} \\
\hline
HAD & A & 0.016 & 0.012 & 0.168 & 92.9 & 74.7 & 0.062 & 0.178 & 0.208 \\
& B & 0.017 & 0.013 & 0.171 & 92.9 & 74.7 & 0.062 & 0.179 & 0.205 \\
& C & 0.015 & 0.011 & 0.160 & 90.9 & 72.7 & 0.064 & 0.179 & 0.212 \\
& D & 0.015 & 0.011 & 0.162 & 90.9 & 74.7 & 0.063 & 0.176 & 0.209 \\
& E & 0.014 & 0.010 & 0.154 & 90.9 & 69.7 & 0.060 & 0.171 & 0.195 \\
& \textit{F} & \textit{0.055} & \textit{0.072} & \textit{0.356} & \textit{99.0} & \textit{84.8} & \textit{0.110} & \textit{0.323} & \textit{0.229} \\
& \textit{G} & \textit{0.081} & \textit{0.071} & \textit{0.390} & \textit{94.9} & \textit{75.8} & \textit{0.118} & \textit{0.389} & \textit{0.259} \\
& \textit{H} & \textit{0.113} & \textit{0.059} & \textit{0.415} & \textit{82.8} & \textit{59.6} & \textit{0.168} & \textit{0.511} & \textit{0.304}
\\[6pt]
CRU & A & 0.032 & 0.025 & 0.238 & 91.9 & 73.7 & 0.084 & 0.251 & 0.245 \\
& B & 0.031 & 0.025 & 0.235 & 91.9 & 75.8 & 0.081 & 0.245 & 0.237 \\
& C & 0.033 & 0.024 & 0.238 & 91.9 & 71.7 & 0.090 & 0.258 & 0.252 \\
& D & 0.032 & 0.023 & 0.234 & 90.9 & 70.7 & 0.087 & 0.255 & 0.245 \\
& E & 0.031 & 0.024 & 0.235 & 91.9 & 73.7 & 0.085 & 0.250 & 0.242 \\
& \textit{F} & \textit{0.089} & \textit{0.097} & \textit{0.432} & \textit{97.0} & \textit{78.8} & \textit{0.131} & \textit{0.416} & \textit{0.274} \\
& \textit{G} & \textit{0.120} & \textit{0.095} & \textit{0.464} & \textit{90.9} & \textit{75.8} & \textit{0.150} & \textit{0.482} & \textit{0.303} \\
& \textit{H} & \textit{0.148} & \textit{0.080} & \textit{0.477} & \textit{84.8} & \textit{62.6} & \textit{0.206} & \textit{0.570} & \textit{0.335} \\
\hline
\end{tabular*}
\tabnotetext[]{tt4}{RMSE: Root Mean Square Error;
$\mathrm{ECP}_{\beta}$: Empirical Coverage Probability at $\beta\%$ confidence level;
$\mathrm{IS}_{\beta}$: Interval Score at $\beta\%$ confidence level;
CPRS: Continuous Ranked Probability Score.}
\tabnotetext[]{tt4}{$^*$ HAD and CRU refer to the two instrumental data sets, with HAD
including the oceans.}
\end{table}

Table~\ref{validateHAD} summarizes the quantitative assessments of the
reconstructions for both the HAD and CRU data sets. The benefit of the
external forcings are readily apparent [cf. \citet{boli1}], as their
inclusion substantially reduces the squared bias, variance and,
consequently, the RMSE, as well as the IS and CRPS (compare Scenario \textup{A}
to \textup{F}, Scenario \textup{D} to \textup{G}, and Scenario \textup{E} to \textup{H}). This corroborates the fact
that the posterior distributions of the coefficients for both the
volcanic and green house gas forcing series are significant.
Moreover, the widths of the 95\% credible intervals are likewise
narrower when the external forcings are included (see Figure~\ref
{combHAD}, below, and Figure~B.16 in the Online Supplement).

When external forcings are included in the reconstruction, the squared
biases, variances and RMSEs are generally similar across the different
error models for each of the two data sets. For the HAD data set, and
among reconstructions that include forcings, Scenarios \textup{A} and \textup{B} are
optimal in terms of ECP; Scenario \textup{E} in terms of CRPS; and there are no
appreciable differences in IS. Note that Scenario~\textup{E} exhibits the worst
ECP, indicating an underestimation of uncertainty compared to Scenarios
\textup{A} and \textup{B}. This is consistent with the rejection of no memory models in
our tests in Section~\ref{Model}. For the CRU data set, Scenario \textup{B} is
optimal in terms of ECP and CRPS, and again there is no appreciable
difference in terms of IS. Based on these validation measures, while
there continues to be support for memory models, there is no clear
indication of a single, best model for the error structures among the
reconstructions that include forcings, with Scenarios \textup{A} and \textup{B} featuring
comparable performance metrics. Indeed, tests for selecting between
long- and short-memory models for climate time series are often
inconclusive [e.g., \citet{percival2001interpretation}].

When forcings are not included, the greater variability of validation
metrics across the scenarios allows for more meaningful ranking of the
error correlation assumptions. For both data sets, Scenario \textup{F} is
optimal in terms of squared bias, RMSE, IS and CRPS.
For the HAD data set, Scenario \textup{G} is optimal in terms of ECP at the 95\%
level but is equally distant from the nominal level as Scenario \textup{F} at
the 80\% level, while for the CRU data set, ECP favors Scenario \textup{F}. In
general, the results indicate that when forcings are not included the
long-memory models play an important role in capturing the correlation
structure in proxies and temperature and should be employed in the
hierarchical model. As discussed in \citet{boli1}, reconstructions are
improved when information is included at a broad range of frequency
scales. In the absence of forcings, which feature long-range
correlations and low-frequency behavior, the inclusion of more highly
structured noise processes leads to marked improvements in the reconstructions.

\subsection{Temperature reconstruction results}\label{TRresult}

\begin{figure}

\includegraphics{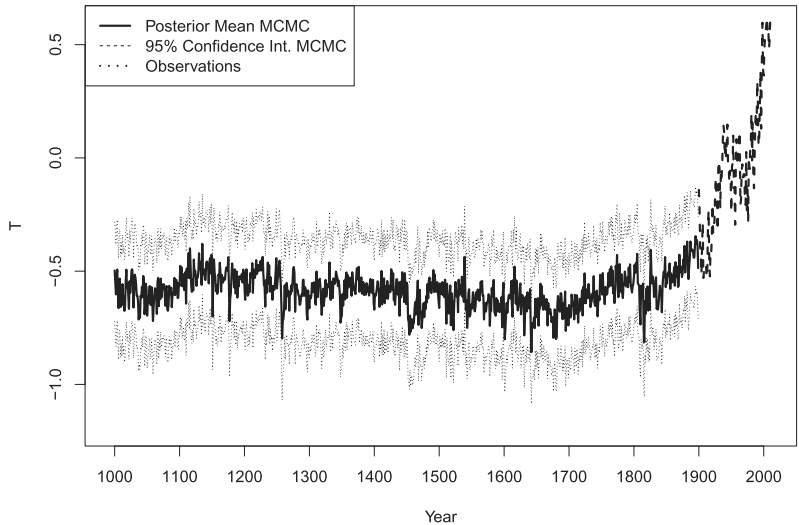}

\caption{Temperature reconstruction (1000--1899) using the HAD data
set, Scenario \textup{B}.}
\label{temprecHADa}
\end{figure}

According to validation measures in Table~\ref{validateHAD}, the
reconstruction scenarios that include forcings are similar to one
another. Here we focus on Scenario B due to slightly better validation
measures for both HAD and CRU data sets. Figure~\ref{temprecHADa}
shows the Scenario B temperature reconstruction together with 95\%
point-wise credible intervals, using the HAD data set. The
reconstruction shows a slight downward trend during the period
1000--1899 [cf. \citet{kaufman2009recent}], and no maxima in the
posterior distributions exceed the levels observed after approximately
1950. The reconstruction for the CRU data set (see Figure~B.14) is qualitatively similar, but features higher variance
due to the more variable CRU temperatures.

%

%
\begin{figure}

\includegraphics{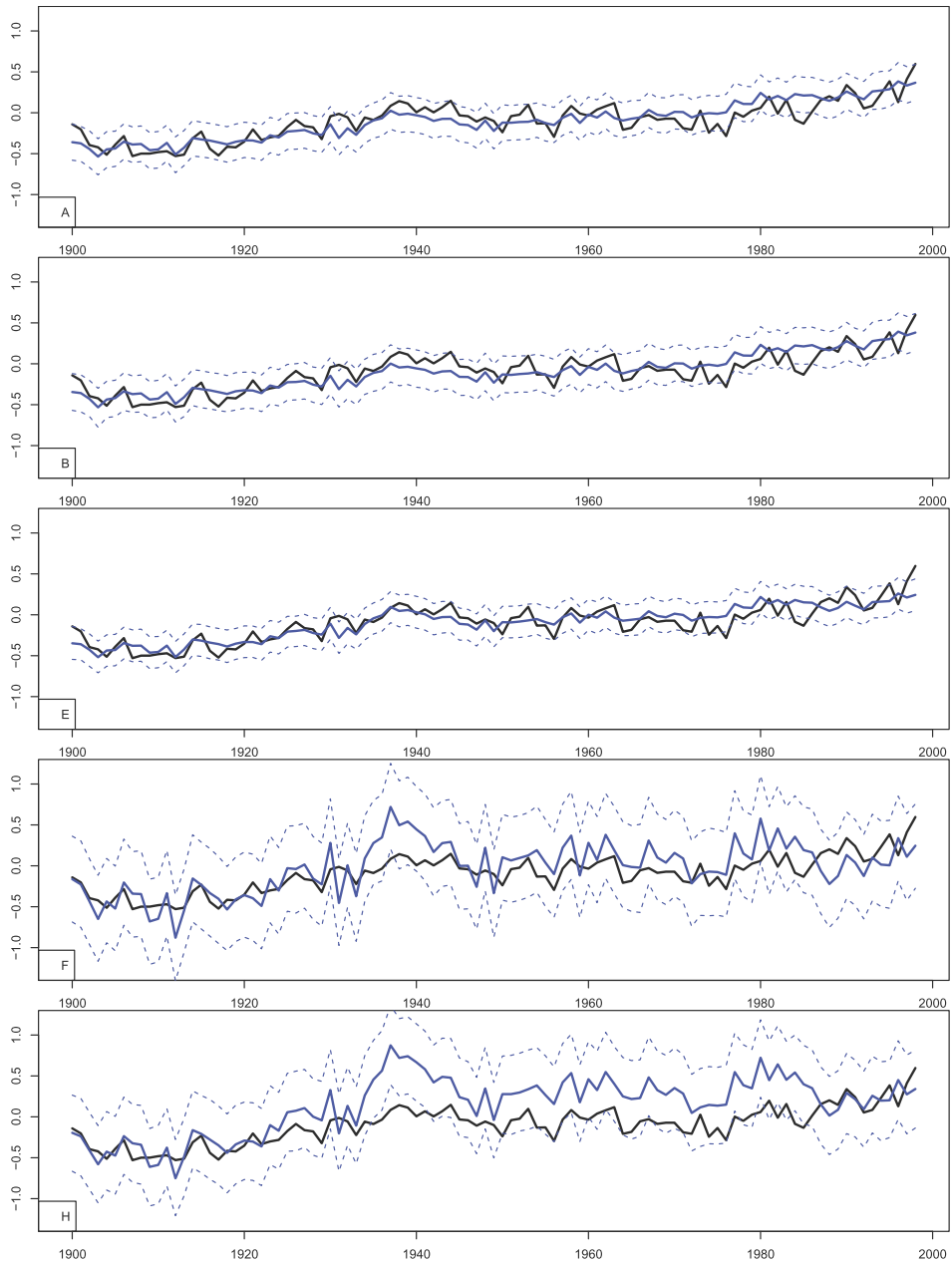}

\caption{Temperature reconstruction (1900--1998) using the HAD data set
under Scenarios \textup{A}, \textup{B}, \textup{E}, \textup{F} and \textup{H}. Black: Observations; Dashed blue:
95\% credible intervals MCMC; Solid blue: posterior mean MCMC.}
\label{temprecHADb}
\end{figure}

In order to evaluate our reconstruction, we use 1900--1998 as an
in-sample validation period. Due to the limited number of available
observations and the necessity of inferring the memory parameters,
out-of-sample validation was not feasible. Figure~\ref{temprecHADb}
shows the posterior mean and 95\% point-wise credible intervals for
predictions using the HAD data in Scenarios A, B, E, F and H, as well
as the actual HAD observations. The scenarios that include forcings (A,
B, E) result in reconstructions that are qualitatively similar to one
another and feature good qualitative agreement with the observations,
with Scenario E exhibiting slightly narrower credible intervals.
Reconstructions resulting from scenarios that exclude the forcings (F
and H) feature greater divergence from the observations---particularly
for Scenario H, which models the error structure as white noise.
Results are similar for the CRU data set (see Figure~B.15). Note that the reduced variability of the posterior mean
as compared with the observations is akin to the predictions from a
linear regression being less variable than the observations. A key
advantage of a Bayesian analysis, such as that used here, is that,
provided the process-level model assumptions are reasonable, the
temporal variability of individual posterior draws will be similar to
that of the actual climate, even while variability of the mean across
them is attenuated [see Figure~2 of \citet{tingley2} for further
discussion]. Repeating the reconstructions with the single lacustrine
record excluded from the reduced proxy leads to similar results; see
Figure~B.17.

\subsection{MCMC diagnostics}

To establish convergence of the MCMC samples, we examine trace plots
(Figures~B.6--B.13) and calculate the
potential scale reduction factor [PSRF; \citet{gelman1992inference}]
and its multivariate version [\citet{brooks1998general}]; see
\citet{brooks} and \citet{cowles} for further details. We
present diagnostic
results for Scenario \textup{A}, as it represents the most complex model for estimation.
If the PSRF is close to unity for all parameters, then the Markov chain
simulation is close to its stationary distribution, while a large PSRF
indicates that the chain has not converged [\citet{gelman1992inference}].
\citet{brooks1998general} provide a generalization that allows for the
computation of a single PSRF for all model parameters.

%
\begin{table}
\tabcolsep=0pt
\caption{Individual and multivariate potential scale reduction factors (PSRF) with the
95\% upper bounds (UB) for individual PSRFs}\label{PSRF}%
\begin{tabular*}{\tablewidth}{@{\extracolsep{\fill}}@{}lccccccccccc@{}}
\hline
& $\bolds{\alpha_{0}}$ & $\bolds{\alpha_{1}}$ & $\bolds{\beta_{0}}$ & $\bolds{\beta_{1}}$ & $\bolds{\beta_{2}}$
& $\bolds{\beta_{3}}$  & $\bolds{\sigma_P^{2}}$ & $\bolds{\sigma_T^{2}}$ & $\bolds{H}$ & $\bolds{K}$ & \textbf{Mul.}\\
\hline
HAD PSRF & 1.01 & 1.01 & 1.01& 1.01& 1.01& 1.00& 1.01& 1.00& 1.01& 1.01 & 1.02\\
UB &1.02 & 1.04 & 1.01& 1.03& 1.02& 1.00& 1.02& 1.01& 1.01& 1.03 & --
\\[3pt]
CRU PSRF &1.00 & 1.01 & 1.03& 1.04& 1.01& 1.01& 1.00& 1.01& 1.00& 1.00 & 1.05\\
UB &1.00 & 1.03 & 1.09& 1.11& 1.01& 1.02 & 1.01& 1.01& 1.00& 1.01& --\\
\hline
\end{tabular*}
\end{table}

For both the HAD and CRU data sets, we run five MCMC simulations, each
of length 5000, and discard the first 1000 samples to allow the chain
to burn in. We compute PSRFs for the scalar parameters of the model
($\alpha_{0}$, $\alpha_{1}$, $\beta_{0}$, $\beta_{1}$, $\beta_{2}$, $\beta
_{3}$, $\sigma_{1}^{2}$, $\sigma_{2}^{2},H,K$) and the multivariate PSRF,
along with their upper 95\% confidence bounds, using the \texttt{coda}
R-package [\citet{coda}]. Results in Table~\ref{PSRF} show that all
the individual PSRFs are relatively close to unity, indicating their
successful convergence to the stationary distribution. The multivariate
PSRF likewise indicates convergence.

\section{Comparison with other works}\label{Comparisons}

\subsection{Comparison with previous reconstructions}
We compare our reconstructions to those reported in \citet{mannzhang},
as both use similar proxy and temperature data sets. \citet{mannzhang}
assume no memory in the error processes, do not include the external
forcings, and present reconstructions, along with uncertainty bands,
based on two regression approaches: composite plus scale (CPS) and
errors in variables (EIV).
The CPS approach computes a weighted average of the proxy data, and
then calibrates this weighted average by matching its mean and variance
to those of the instrumental temperature data during their overlap
period. The EIV regression approach allows for errors in both the
dependent and independent variables, and we refer to
Mann et~al. (\citeyear{mannzhang,Mannetal2008supp}) for details. The EIV and CPS
reconstructions, and their associated uncertainty estimates, are
available online\footnote{\url{http://www.ncdc.noaa.gov/paleo/pubs/mann2008/mann2008.html}.} as
decadally smoothed time series, as \citet{mannzhang} focuses on
low-frequency climate variability.
In contrast, the reconstructions we present here are available at
annual temporal resolution, with no smoothing.
In comparisons, we show the posterior mean and uncertainty of our
reconstructions at annual resolution, and additionally include the
posterior mean that results from first smoothing each posterior draw
with a Butterworth filter\footnote{Our calculations are based on the
Matlab code associated with \citet{mannzhang}, posted online at
\url{http://www.ncdc.noaa.gov/paleo/pubs/mann2008/mann2008.html}. We smooth
using the \mbox{\texttt{filtfilt}} command in the R package ``signal.''} with
cutoff frequency equal to 0.1 cycles${}/{}$year.

\begin{figure}

\includegraphics{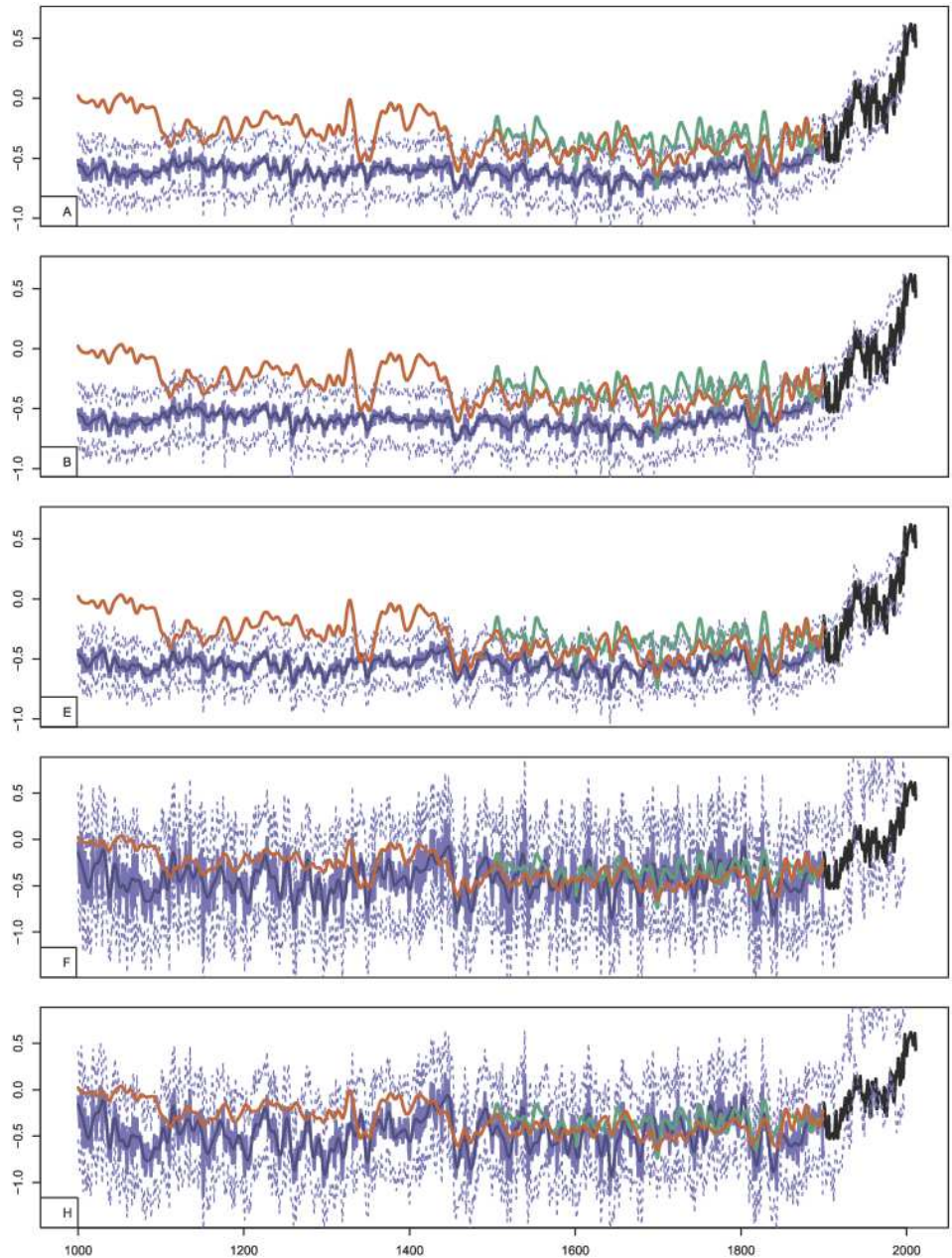}

\caption{Comparisons between Scenarios \textup{A}, \textup{B}, \textup{E}, \textup{F}, \textup{H} and CPS and EIV
reconstructions in \citet{mannzhang} using the HAD data set. Black:
Observations; Purple: posterior mean reconstruction with 95\% credible
intervals; Orange: EIV; Green: CPS; Dark Purple line: mean of smoothed
posteriors.}\label{combHAD}
\end{figure}

Figure~\ref{combHAD} compares our reconstructions using the HAD data,
and under Scenarios \textup{A}, \textup{B}, \textup{E}, \textup{F} and \textup{H}, to those from \citet{mannzhang}.
In all cases, and especially when including the forcings, our
reconstructions are generally cooler than both the EIV and CPS
reconstructions from \citet{mannzhang}, particularly during the
1000--1400 interval, and feature a smaller amplitude of
pre-instrumental temperature variability. We are not the first to
report a lower variability than \citet{mannzhang}---for example,
\citet{consortiumcontinental2013} report a change in 30 year average
temperatures between 1000~AD and the 1800s of about $0.3^{\circ}$C,
compared with about $0.5^{\circ}$C for \citet{mannzhang}; see Figure~4
of \citet{consortiumcontinental2013}.

The model settings of \citet{mannzhang} are most similar to our
Scenario \textup{H}, which includes neither the forcings nor the long-memory
processes. Indeed, the EIV predictions from \citet{mannzhang} are
visually most similar to smoothed Scenario \textup{H} results, and 88.4\% of the
EIV predictions from \citet{mannzhang} fall within the 95\% point-wise
credible intervals for the smoothed Scenario \textup{H} results. Results are
similar when using the CRU data set (Figure~B.16).

%

To facilitate numerical comparisons with the \citet{mannzhang}
reconstruction, we recalculate validation metrics for Scenario H after
first smoothing each posterior draw; results are shown in Table~\ref
{comparisonMann} for the 20th century validation interval.
The main difference between our smoothed Scenario H results and the
\citet{mannzhang} results is in terms of squared bias, with the
\citet{mannzhang} reconstruction featuring biases that are about
an order of
magnitude smaller and variances that are about 1.5--2 times larger. The
net result is that the \citet{mannzhang} reconstructions feature
smaller RMSE than our smoothed Scenario H, on par with results from our
annually resolved Scenarios A, B, C, D and E.
As measured by the ECP, the uncertainties for the \citet{mannzhang}
reconstructions are too wide, in the sense that the empirical coverage
rate is greater than the nominal rate.
The uncertainties for our smoothed Scenario H are smaller than that in
\citet{mannzhang}, but due to the relatively large bias, the ECPs
appear to be too low compared to their nominal value. The Interval
Scores for the smoothed \citet{mannzhang} reconstructions are much
better than those for our smoothed Scenario H and, like the RMSE, are
similar to those for our annually resolved Scenarios A, B, C, D and E
which carry small squared bias by including the forcings (see
Table~\ref{validateHAD}).

%
\begin{table}
\tabcolsep=0pt
\caption{Comparison between Scenarios \textup{H} and CPS and EIV
reconstructions in \protect\citet{mannzhang}\protect\tabnoteref{tt6}}\label{comparisonMann}
\begin{tabular*}{\tablewidth}{@{\extracolsep{\fill}}@{}lccccd{3.1}ccc@{}}
\hline
& \textbf{Scenarios} & \textbf{Sq. bias} & \textbf{Variance} & \textbf{RMSE} & \multicolumn{1}{c}{\textbf{ECP}$\bolds{_{95}}$} & \textbf{ECP}$\bolds{_{80}}$ & \textbf{IS}$\bolds{_{95}}$
& \textbf{IS}$\bolds{_{80}}$ \\
\hline
HAD & H (smoothed) &0.100 &0.012 &0.335 & 41.4 & 33.3 & 0.46 & 0.71 \\
& CPS &0.009 &0.024 &0.183 & 100.0 & 96.9 & 0.06 & 0.16 \\
& EIV &0.003 &0.022 &0.157 & 99.0 & 99.0 & 0.06 & 0.23
\\[3pt]
CRU & H (smoothed) &0.121 &0.016 &0.371 & 48.5 & 36.4 & 0.45 & 0.73 \\
& CPS &0.017 &0.025 &0.207 & 99.0 & 99.0 & 0.07 & 0.25 \\
& EIV &0.006 &0.021 &0.163 & 98.0 & 98.0 & 0.07 & 0.17 \\
\hline
\end{tabular*}
\tabnotetext{tt6}{The statistics for EIV and CPS reconstructions are calculated using
the estimated standard deviations associated with \citet{mannzhang}.
They\vspace*{1pt} are posted as ``2-sigma uncertainties'' ($S$), hence, the formula
for their 95\% confidence bands is $M_t\pm\frac{1.96}{2}S$, where
$M_t$ is their predicted temperature mean.}
\end{table}

We caution against drawing substantive conclusions from the comparison
of the validation and scoring metrics between the \citet{mannzhang}
results and the smoothed Scenario H, as numerous lines of evidence
indicate that Scenario H is the least appropriate of the eight
scenarios explored here: validation metrics and scores (Table~\ref
{validateHAD}) are generally worst for Scenario H; the inclusion of the
forcings is motivated by the scientific understanding of their
connection with temperatures; and the inclusion of the long-memory
processes in the absence of forcings is driven by the structure of the
data. Indeed, we view Scenario H as a misspecified model, and the high
squared bias and associated inadequacies of the ECPs are therefore to
be expected. Perhaps the most telling conclusion to be drawn from the
numerical comparisons is that our annually resolved Scenarios A, B, C
and D, which include the forcings as well as short- and/or long-memory
processes, are comparable in terms of RMSE and Interval Scores to the
decadally resolved \citet{mannzhang} results while featuring ECPs which
are closer to their nominal values.

Finally, we note that the proxy selection and modeling treatments do
differ between our Scenario H and the reconstructions in \citet
{mannzhang} so that the comparison remains imperfect. In particular, we
note that the \citet{mannzhang} reconstructions include proxies with
decadal resolution, whereas here we focus on proxies with annual
resolution. Indeed, the CPS reconstruction is performed after smoothing
all proxies to a common decadal resolution, while the EIV
reconstruction is based on a ``hybrid'' frequency approach that involves
separate calibrations to infer climate on interdecadal (periods longer
than 20 years) and interannual (periods shorter than 20 years)
timescales [Mann et~al. (\citeyear{mann2007robust,Mannetal2008supp})]. Due to the
differing methods and the focus on lower frequency variability in
\citet{mannzhang}, the differing validation metrics between our smoothed
Scenario H and those for the \citet{mannzhang} reconstructions are not
surprising.

\subsection{Transient climate response}

The Fourth Assessment Report of the IPCC [see page 723 in \citet
{ipcc9}] refers to the ``transient climate response'' (TCR) as the
``global mean temperature change that is realized at the time of CO$_2$
doubling $\ldots$ TCR is therefore indicative of the temperature trend
associated with external forcing, and can be constrained by an
observable quantity, the observed warming trend that is attributable
to greenhouse gas (GHG) forcing.'' In our model, the transient response
to a doubling of GHG is functionally related to the parameter~$\beta
_3$, and the resulting estimates of TCR are based on the instrumental
temperature record since 1900 and proxy and forcing information over
the past millennium.
We believe that our Bayesian approach to computing the transient
response to GHG forcing from both instrumental and proxy observations,
without recourse to global climate models, is new to the field.

Taking into account the transformations applied to the CO$_2$ series,
we define TCR in terms of $\beta_3$ as
\begin{eqnarray*}
\mathrm{TCR}&:=&\beta_3{\log2}/{\sigma(\log\mathbf{C})},
\end{eqnarray*}
where $\sigma(\log\mathbf{C})$ is the standard deviation of the
logarithm of the GHG series $\mathbf C$ and is computed over the entire
period 1000--1998.
An important advantage of Bayesian estimation is the possibility of
obtaining a sample estimate of the marginal posterior distribution of
$\beta_3$ given the data, from which we can compute a nonparametric
estimator of the probability density function for TCR that accounts for
the uncertainties in all other parameters in the model.

%
\begin{figure}

\includegraphics{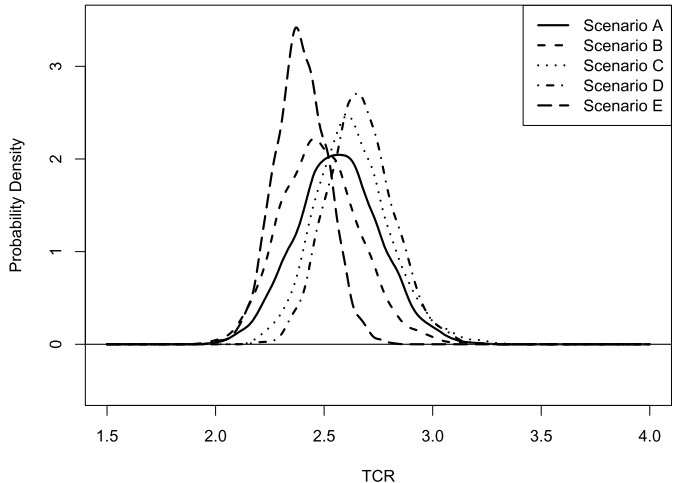}

\caption{Estimates of probability density function for Transient
Climate Response (TCR) in degree Celsius for Scenarios \textup{A}, \textup{B}, \textup{C}, \textup{D} and \textup{E} (HAD).}
\label{figsensibility}
\end{figure}

We present results of TCR estimates using the global land and marine
HAD data set, for the five scenarios that include the forcings:
Scenarios A, B, C, D and E (Figure~\ref{figsensibility}).
There is substantial variability between the TCR estimates from the
five scenarios. TCR estimates are the lowest and most sharply peaked
for the memory-free Scenario E, with a median around $2.39^{\circ}$
and an approximate 95\% credible interval of $[2.16, 2.63]^{\circ}$C.
The TCR distributions become progressively broader as more memory is
included, in Scenarios D, then in B and C, and finally in the fully
long-memory Scenario A which features the broadest 95\% credible
interval of $[2.19, 2.95]^{\circ}$C. A quantitative explanation for
this increasing uncertainty behavior can be found by inspecting the
formula for the covariance matrix $\Omega_{\beta}$ of the posterior
distribution of the vector $\beta$ given $T$: from formula (A.4) in the Online Supplement~\textup{A} [see
\citet{Barboza2014}], one sees that $\Omega_{\beta}$ is the
inverse of a
matrix which is affine in $\Sigma_K^{-1}$, that is, affine in the
inverse of the covariance matrix for the noise model being used in each
scenario. It is known [see, e.g., \citet{Palma2003}\footnote
{In this paper, the authors provide the estimate $\lambda_{n,n} \asymp
n^{2H-1}$ for the top eigenvalue of the covariance matrix of a vector
of $n$ contiguous terms of a stationary sequence whose covariance
matrix $\rho$ satisfies $\rho( n ) \sim cn^{2H-2}$, which is the
case for our fGn sequence. Thus, indeed, $\lambda_{n,n}$ is roughly
increasing in $H$ for all $H\in(0.5,1)$. \citet{Palma2003} state this
result in the case of the ARFIMA process, in Example~2 on pages 99--100,
but an inspection of their proof shows that the result holds for all
$\rho$ satisfying the above asymptotics.}]
that the magnitude (e.g., the operator norm) of $\Sigma_K$ increases
with memory length; this and the formula for $\Omega_{\beta}$ can
explain the increasing behavior we observe in Figure~\ref{figsensibility}.

On the other hand, the progression of posterior medians for the TCRs is
not monotone. Scenario D [AR(1) errors] features the largest median TCR
value: $2.66^{\circ}$C, followed by Scenario C with short memory in
the proxy model: $2.62^{\circ}$C. The two scenarios with long memory
in the proxy model, A and B, have lower median TCR values, respectively,
$2.56^{\circ}$C and $2.47^{\circ}$C, with Scenario E (no memory)
having the lowest median of $2.39^{\circ}$C, as reported above. The
formula for the posterior distribution of the vector $\beta$ given $T$
is again helpful: the posterior mean is the product of the increasing
$\Omega_{\beta}$, as discussed above, and of a matrix $\Delta_{\beta
}$ which is affine in $\Sigma_K^{-1}$, thus with presumably decreasing
magnitude with respect to memory length; the competition between these
two effects could induce nonmonotonicity with respect to memory length.

To arrive at a best estimate of TCR, we mix with equal weights the
posterior estimates from Scenarios A and B, yielding a median TCR of
about $2.5^{\circ}$C with a combined $95\%$ credible interval of about
$[2.16,2.92]^{\circ}$C.
Scenarios A and B feature superior validation metrics as compared with
Scenarios C and D, while Scenario E presumably under-reports TCR
uncertainty because it is based on a model that lacks memory. We
therefore focus on TCR estimates derived from an equally weighted
mixture of estimates from Scenarios A and B. It is not possible to
select between these two scenarios using model diagnostics and, as both
include memory, our estimates are conservative with respect to uncertainty.

It is instructive to compare our five TCR distributions reported here
with the consensus (expert assessment) recently released in the IPCC's
Fifth Assessment Report [\citet{ipcc5th10,ipcc5th12}], where TRC is
reported as ``likely'' within the interval $[1,2.5]^{\circ}$C and
``extremely unlikely'' to exceed $3.0^{\circ}$C. Using the IPCC's
definitions/guidance on uncertainty language, these expert assessments
can be interpreted as meaning that the probability of the estimated TCR
from one of our scenarios falling within the $[1,2.5]^{\circ}$C
interval should exceed 0.66, while the probability that the estimated
TCR exceeds $3.0^{\circ}$C should not exceed $0.05$. For all scenarios
reported in Figure~\ref{figsensibility}, the posterior probability
that the TCR exceeds $3.0^{\circ}$C is in all cases lower than $0.05$.
This exceedance probability is essentially zero for Scenario E, which
features the narrowest TCR distribution: Scenario~E presumably
underestimates uncertainty by using no memory for modeling errors. As
for falling in the interval $[1,2.5]^{\circ}$C with probability around
0.66, Scenario E does satisfy this condition; Scenario B nearly does;
for Scenario A the probability is closer to 0.5; but Scenarios C and D,
with their significantly higher median values, fail the condition by
some margin. Our best estimate, derived from mixing Scenarios A~and~B,
meets the TCR upper bound condition: the probability that it exceeds
$3.0^\circ$C is about $0.011$. It falls slightly short of meeting the
confidence interval condition: the probability that it falls within the
interval $[1,2.5]^\circ$C is about $0.47$.

All of our reported TCRs are on the high side compared to the latest
IPCC consensus, and as compared with several specific recent studies
which have arrived at TCR estimates by combining information from
models and the instrumental temperature record.
\citet{Gillett2012p8513} produce a TCR estimate of 1.3--1.8$^{\circ}$C
using the global HAD data set and a single global climate model, but
note that this TCR estimate may be unrealistically narrow as it results
from a single climate model. A more recent study [\citet
{gillett2013constraining}] that combines information from an ensemble
of models and the instrumental record results in a wider range of TCR
estimates, 0.9--2.3$^{\circ}$C, featuring greater overlap with our
results. \citet{Otto2013p8514} use global, decadal averages of the HAD
data set over the 1970--2009 to arrive at a data-based TCR estimate in
the range of 0.7--2.5$^{\circ}$C, but caution against strong
conclusions based on a such a short time interval. A particularly high
estimate of TCR, of at least 2.5--3.6$^{\circ}$C, is reported by
\citet{tung2008}, based on an analysis of the 11-year solar cycle.

Hence, both the specific model-data fusion studies discussed in the
previous paragraph and in \citet{ipcc5th10}, as well as the synthesis
provided by the IPCC Fifth Assessment Report, generally feature broader
uncertainties and are peaked at lower values as compared to our
posterior estimates of TCR. Indeed, only one of the estimated TCR
distributions shown in Figure~10.20 of \citet{ipcc5th10} is peaked
at a
value greater than $2^{\circ}$C, while the high estimate of \citet
{tung2008} is explained as resulting from solar forcing having a
different mechanistic effect on climate [\citet{ipcc5th10}].
Interestingly, the single plotted TCR distribution peaked at greater
than $2^{\circ}$C is that of \citet{harris2013probabilistic}, which
estimates TCR using a Bayesian approach that combine information from
GCMs and recently observed temperature changes. A possible cause for
the narrower uncertainties and higher TCR values estimated here is the
more extensive use of data, in terms of both variety (instrumental
temperatures, proxies, and estimates of CO$_2$, volcanic and solar
forcings) and duration (observations over the last millennium).

\section{Conclusions and discussion}
\label{Conclusions}

We use a comprehensive multiproxy data set to produce new
reconstructions of NH temperature anomaly time series back to 1000 AD
and systematically evaluate the effects of including or excluding
external drivers of climate variability, and of assuming the error
processes feature long, short or no memory, by considering eight
modeling scenarios. Hierarchical Bayesian models are used throughout as
they provide a natural framework for integrating the different
information sources---proxy and instrumental temperatures
observations, and time series of solar, greenhouse gas and volcanic
forcings. Bayesian inference additionally permits for estimation of all
unknown quantities, including past temperatures, and facilitates
uncertainty propagation.

While the possibility of long memory was suggested by exploratory data
analysis, and the significance of long-memory parameters verified by
Bayesian estimation, model diagnostics indicated that short- and
long-memory models yield comparable results provided the climate
forcings are incorporated into the reconstructions. The inclusion of
the external forcings is motivated from physical principles and the
conclusions of \citet{boli1}, and additionally allows for
estimation of
the transient climate response. While our TCR estimates are near the
upper bound of the expert-derived ``extremely likely'' interval provided
in the IPCC Fifth Assessment Report [\citet{ipcc5th10}], they do not
violate this uncertainty consensus, and we note that our estimate is
based on both the instrumental and paleoclimate records, and does not
rely on GCMs.

If the forcings are excluded from the reconstruction, as is necessary
for reconstructions to be suitable for GCM assessment exercises, the
long-memory processes substantially improve the quality of the
reconstructions. The scenario with neither forcings nor memory is
similar to the benchmark reconstruction of \citet{mannzhang},
though we
note that there remain differences in both method and data usage. Our
reconstructions generally indicate cooler temperatures than those of
\citet{mannzhang}, particularly before the year 1400.

The basic framework presented in this paper can be extended in several
directions, and we anticipate that doing so will produce further
insights into the climate of the late Holocene. An obvious extension is
to incorporate a spatial element, by combining the model used here with
the space--time model in \citet{tingley1}. Doing so would require
generalizing the reduced-proxy framework and instead specifying a
separate long-memory error model for each proxy time series, or perhaps
a common model for each proxy type [cf. \citet{tingley1}]. Such an
implementation would pose technical challenges, as the estimation of
the long-memory parameters is the most numerically demanding component
of the analysis. Prior scientific understanding of the mechanisms by
which the proxies record variations in the climate may be helpful in
selecting appropriate temporal correlation models for the residuals,
and can potentially be used to simplify calculations.
Such a computationally demanding generalization may be a more
scientifically defensible use of the proxies and may allow for further
insights into the proxy--climate relationship.

\section*{Acknowledgments}
The authors thank the Editor, three anonymous referees, and Michael Stein
for their helpful comments and suggestions.

\begin{supplement}
\stitle{Supplement to: ``Reconstructing past temperatures from natural proxies and estimated climate forcings using short- and long-memory models''}
\slink[doi]{10.1214/14-AOAS785SUPP} 
\sdatatype{.pdf}
\sfilename{10.1214/14-AOAS785\_supp.pdf}
\sdescription{We provide a background on long-memory models, the
multitaper estimator and scoring rules together with some calculations
of our model's posterior distributions. Finally, we include additional
plots and tables.}
\end{supplement}


%

\printaddresses
\end{document}